\documentclass[10pt]{article}
\usepackage[OE]{express}

\usepackage{graphicx}
\usepackage{float,comment}
\usepackage[tight,footnotesize]{subfigure}
\usepackage{mathdots}
\usepackage{enumerate}
\usepackage{cite}
\usepackage{amsmath,amssymb,bm}

\def\rbf{{\mathbf{r}}}

\def\ubf{{\mathbf{u}}}
\def\xbf{{\mathbf{x}}}

\def\ybm{{\bm{i}}}
\def\ybm{{\bm{y}}}
\def\cbm{{\bm{c}}}
\def\Wbm{{\bm{W}}}
\def\Abm{{\bm{A}}}

\def\drm{{\mathrm{d}}}
\def\argmin{\mathop{\mathrm{arg\,min}}}

\def\mum{{\mu\mathrm{m}}}
\renewcommand{\eqref}[1]{{Eq.~(\ref{#1})}}

\begin{document}

\title{Multiplexed phase-space imaging for \\ 3D fluorescence microscopy} 

\author{Hsiou-Yuan Liu, Jingshan Zhong and Laura Waller}
% \author{Author One,\authormark{1} Author Two,\authormark{2,*} and Author Three\authormark{2,3}}

\address{Department of Electrical Engineering and Computer Sciences, University of California, Berkeley, USA}
% \address{\authormark{1}Peer Review, Publications Department, The Optical Society, 2010 Massachusetts Avenue NW, Washington, DC 20036, USA\\
% \authormark{2}Publications Department, The Optical Society, 2010 Massachusetts Avenue NW, Washington, DC 20036, USA\\
% \authormark{3}Currently with the Department of Electronic Journals, The Optical Society, 2010 Massachusetts Avenue NW, Washington, DC 20036, USA}

%\email{\authormark{1}hyliu@eecs.berkeley.edu}
\email{\authormark{*}lwaller@alum.mit.edu} %% email address is required


\begin{abstract}
Optical phase-space functions describe spatial and angular information simultaneously; examples include light fields in ray optics and Wigner functions in wave optics. Measurement of phase-space enables digital refocusing, aberration removal and 3D reconstruction. High-resolution capture of 4D phase-space datasets is, however, challenging. Previous scanning approaches are slow, light inefficient and do not achieve diffraction-limited resolution. Here, we propose a multiplexed method that solves these problems. We use a spatial light modulator (SLM) in the pupil plane of a microscope in order to sequentially pattern multiplexed coded apertures while capturing images in real space. Then, we reconstruct the 3D fluorescence distribution of our sample by solving an inverse problem via regularized least squares with a proximal accelerated gradient descent solver. We experimentally reconstruct a 101 Megavoxel 3D volume (1010$\times$510$\times$500$\mum$ with NA 0.4), demonstrating improved acquisition time, light throughput and resolution compared to scanning aperture methods. Our flexible patterning scheme further allows sparsity in the sample to be exploited for reduced data capture.
\end{abstract}

 % REPLACE WITH CORRECT OCIS CODES FOR YOUR ARTICLE, MINIMUM OF TWO; Avoid using the OCIS codes for “General” or “General science” whenever possible.
%For a complete list of OCIS codes, visit: https://www.osapublishing.org/oe/submit/ocis/

%%%%%%%%%%%%%%%%%%%%%%% References %%%%%%%%%%%%%%%%%%%%%%%%%
% \begin{thebibliography}{99}
% 
% \bibitem{gallo99} K. Gallo and G. Assanto, ``All-optical diode based on second-harmonic generation in an asymmetric waveguide,'' \josab {\bfseries 16}(2), 267--269 (1999).
% 
% \end{thebibliography}

\bibliography{ref}
\bibliographystyle{osajnl}

%%%%%%%%%%%%%%%%%%%%%%%%%%  body  %%%%%%%%%%%%%%%%%%%%%%%%%%
\section{Introduction}
\label{sec:intro}
3D fluorescence microscopy is a critical tool for bioimaging, since most samples are thick and can be functionally labeled. High-resolution 3D imaging typically uses confocal~\cite{cox1982super}, two-photon~\cite{helmchen2005deep} or light sheet microscopy~\cite{planchon2011rapid}. Because these methods all involve scanning, they are are inherently limited in terms of speed or volume. Light field microscopy~\cite{Levoy2006lfm,broxton2013wave}, on the other hand, achieves single-shot 3D capture, but sacrifices resolution. High resolution \textit{and} single-shot capture are possible with coded aperture microscopy~\cite{Schechner1996DH,pavani2009three,Shechtman2014,ji10adaptive}; however, this requires an extremely sparse sample. Here, we describe a multi-shot coded aperture microscopy method for high-resolution imaging of large and dense volumes with efficient data capture.

Our work fits into the framework of phase-space optics, which is the general term for any space-angle description of light~\cite{Testorf:2009xy,bastiaans78,accardi09,Zhang:2009}. Light fields, for example, are phase-space functions for ray optics with incoherent light. The light field describes each ray's position and angle at a particular plane, which can be used for digital refocusing~\cite{Ng2005}, 3D imaging, aberration correction~\cite{ng2007digital} or imaging in scattering~\cite{pegard2016compressive}. In microscopy, wave-optical effects become prominent and so ray optics no longer suffices if one wishes to achieve diffraction-limited resolution. Hence, we use Wigner functions, which are the wave-optical analog to light fields~\cite{Zhang:2009,accardi09}. 

The Wigner function describes spatial and spatial frequency (propagation angle) information for a wave-field of arbitrary coherence. It converges to the light field in the limit of incoherent ray optics~\cite{Zhang:2009}. Capturing Wigner functions is akin to spatial coherence imaging~\cite{Marks:99,wood2014using}, since they contain the same information as mutual intensity and coherent mode decompositions~\cite{Testorf:2009xy, christodoulides01}. In the case of fluorescence microscopy, where the object is a set of incoherent emitters, Wigner representations define the incoherent 3D Optical Transfer Function (OTF)~\cite{Testorf:2009xy}. 

The 4D nature of phase space (two spatial and two spatial-frequency dimensions) poses significant measurement challenges. Single-shot schemes (e.g. lenslet arrays~\cite{tian2013wigner}) must distribute the 4D function across a 2D sensor, severely restricting resolution. Scanning aperture methods~\cite{waller12NP,Liu15,Cho:12} are slow ($\sim$minutes) and light inefficient. We seek here a flexible trade-off between capture time and resolution, with the ability to exploit sparsity for further data reduction.

\begin{figure}[b]
\centering
\includegraphics[width=0.97\textwidth]{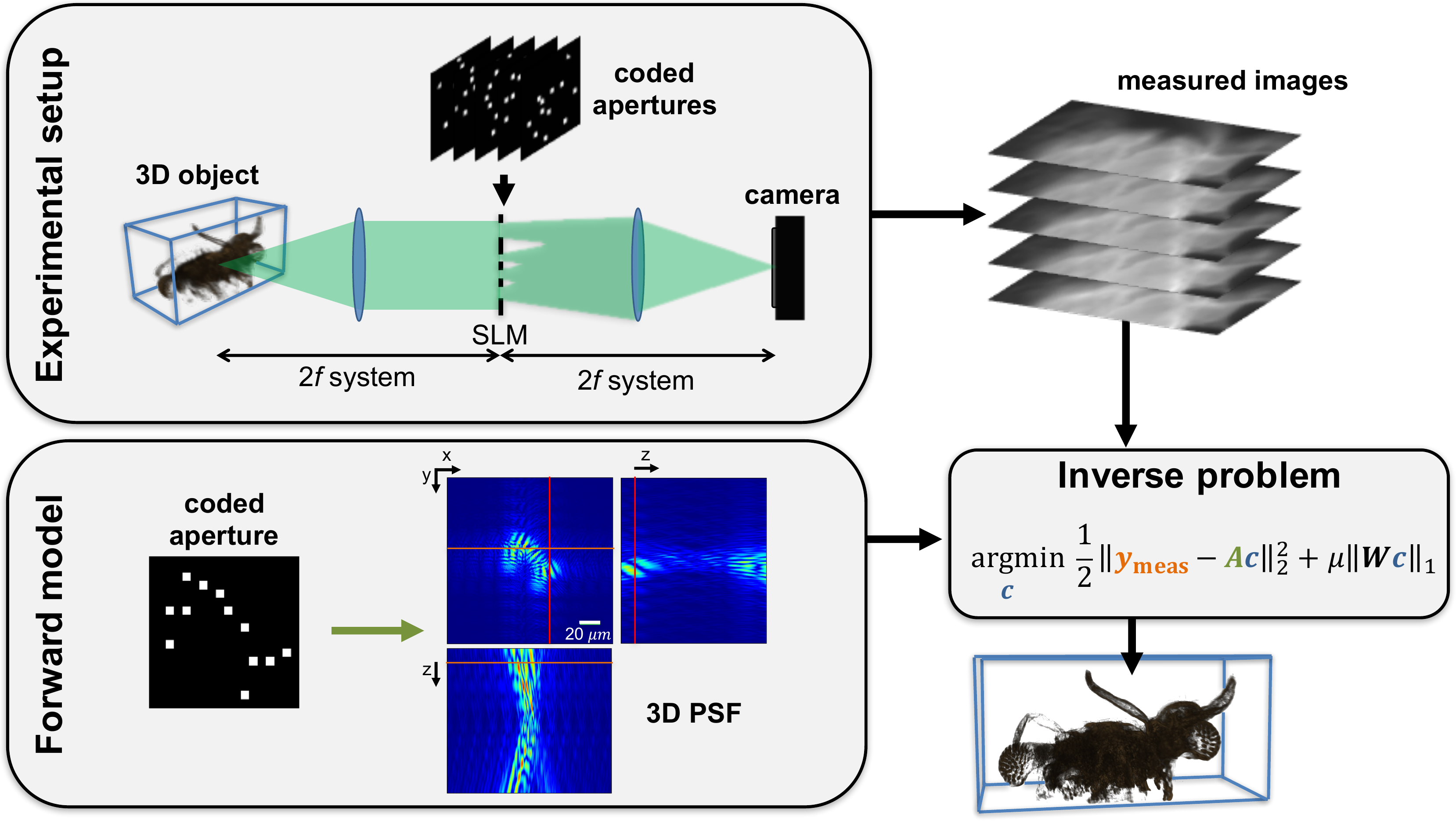}
\caption{\label{fig:scheme-tot}Phase-space multiplexing for 3D fluorescence microscopy. The microscope (a $4f$ system) uses a Spatial Light Modulator (SLM) in Fourier space to implement multiple coded apertures, while capturing 2D intensity images in real space for each. Our wave-optical forward model $A$ relates the object $c$ to the measured images $y_\mathrm{meas}$ for each pattern. The inverse problem recovers the object, subject to sparsity priors where applicable.}
\end{figure}

Our experimental setup consists of a widefield fluorescence microscope with a spatial light modulator (SLM) in Fourier space (the pupil plane). The SLM implements a series of quasi-random coded aperture patterns, while collecting real space images for each (Fig.~\ref{fig:scheme-tot})~\cite{liu20164d}. The recovered 4D phase space has very large pixel count (the product of the pixel counts of the SLM and the sensor) $\sim 10^{12}$. Compared to scanning aperture methods~\cite{waller12NP,Liu15}, the new scheme has three major benefits. \textit{First}, it achieves better resolution by capturing high-frequency interference effects (high-order correlations). This enables diffraction-limited resolution at the microscope's full numerical aperture (NA). \textit{Second}, we achieve higher light throughput by opening up more of the pupil in each capture; this can be traded for shorter exposure time and faster acquisition. \textit{Third}, the multiplexed nature of the measurements means that we can employ compressed sensing approaches (when samples are sparse) in order to capture fewer images without sacrificing resolution. This means that the number of images required scales not with the reconstructed number of resolved voxels, but rather with the sparsity of the volume. 

Our method can be thought of as a multi-shot coded aperture scheme for diffraction-limited 3D fluorescence microscopy. It is analogous to coded aperture photography~\cite{levin2007image,Liang2008,green2007multi,chang2016variable}; however, we use a wave-optical model to account for diffraction effects, so intensity measurements are nonlinear with complex-field. Fluorescent imaging allows a simplification of the forward model, since each fluorophore is spatially coherent with itself but incoherent with all other fluorophores. Our reconstruction algorithm then becomes a large-scale inverse problem akin to multi-image 3D deconvolution, formulated as a convex $\ell_1$ regularized least-square error problem and solved by a fast iterative shrinkage-thresholding algorithm (FISTA)~\cite{beck2009fista}. 

\section{Theory and methods}
We use the Wigner function~(WF) and its close relative, the Mutual Intensity~(MI) function, to describe our multiplexing scheme. The WF describes position and spatial frequency, where spatial frequency can be thought of as the direction of ray propagation in geometrical optics. The concept was introduced by Walther~\cite{WALTHER:68} to describe the connection between ray intensity and wave theory. Here we assume that the light is quasi-monochromatic, temporally stationary and ergodic, so the Wigner function is~\cite{bastiaans78,Testorf:2009xy}:
%\begin{align}
%	W(\ubf,\rbf) 
%	&\triangleq \iint \left\langle E^* (\rbf-\frac{\Delta\rbf}{2}) E (\rbf+\frac{\Delta\rbf}{2}) \right\rangle e^{-i2\pi \ubf\cdot(\Delta\rbf)} \drm^2(\Delta\rbf)\label{eq:wdf-def0}\\
%	&=\iint \left\langle \tilde{E}^* (\ubf-\frac{\Delta\ubf}{2}) \tilde{E} (\ubf+\frac{\Delta\ubf}{2}) \right\rangle e^{i2\pi (\Delta\ubf)\cdot \mathbf{r}} \drm^2(\Delta\ubf),\label{eq:wdf-def}
%\end{align}
\begin{align}
	W(\ubf,\rbf) 
	&\triangleq\iint \left\langle \tilde{E}^* (\ubf-\frac{\Delta\ubf}{2}) \tilde{E} (\ubf+\frac{\Delta\ubf}{2}) \right\rangle e^{i2\pi (\Delta\ubf)\cdot \mathbf{r}} \drm^2(\Delta\ubf),\label{eq:wdf-def}
\end{align}
where $\rbf=(x,y)$ denotes transverse spatial coordinates, $\ubf=(u_x,u_y)$ denotes spatial frequency coordinates and $\langle\cdot\rangle$ denotes ensemble average. The quantity contained by angle brackets,
\begin{equation}\label{eq:mi-def}
\mathrm{MI}(\ubf,\Delta\ubf)\triangleq \left\langle \tilde{E}^* (\ubf-\frac{\Delta\ubf}{2}) \tilde{E} (\ubf+\frac{\Delta\ubf}{2}) \right\rangle,
\end{equation}
is, apart from a coordinate transform, the Mutual Intensity. $E(\mathbf{r})$ is a spatially coherent electric field (e.g. from a single fluorophore) and $\tilde{E}(\ubf)$ is its Fourier transform. The ensemble average allows representation of both coherent and partially (spatially) coherent light. Here, we assume that the object is a 3D volume of incoherent emitters with no occlusions. Thus, the phase-space description of light from the object is a linear sum of that from each fluorophore.

\subsection{General forward model}\label{sec:multiplex}
\begin{figure}[tb]
\centering
\includegraphics[width=0.97\textwidth]{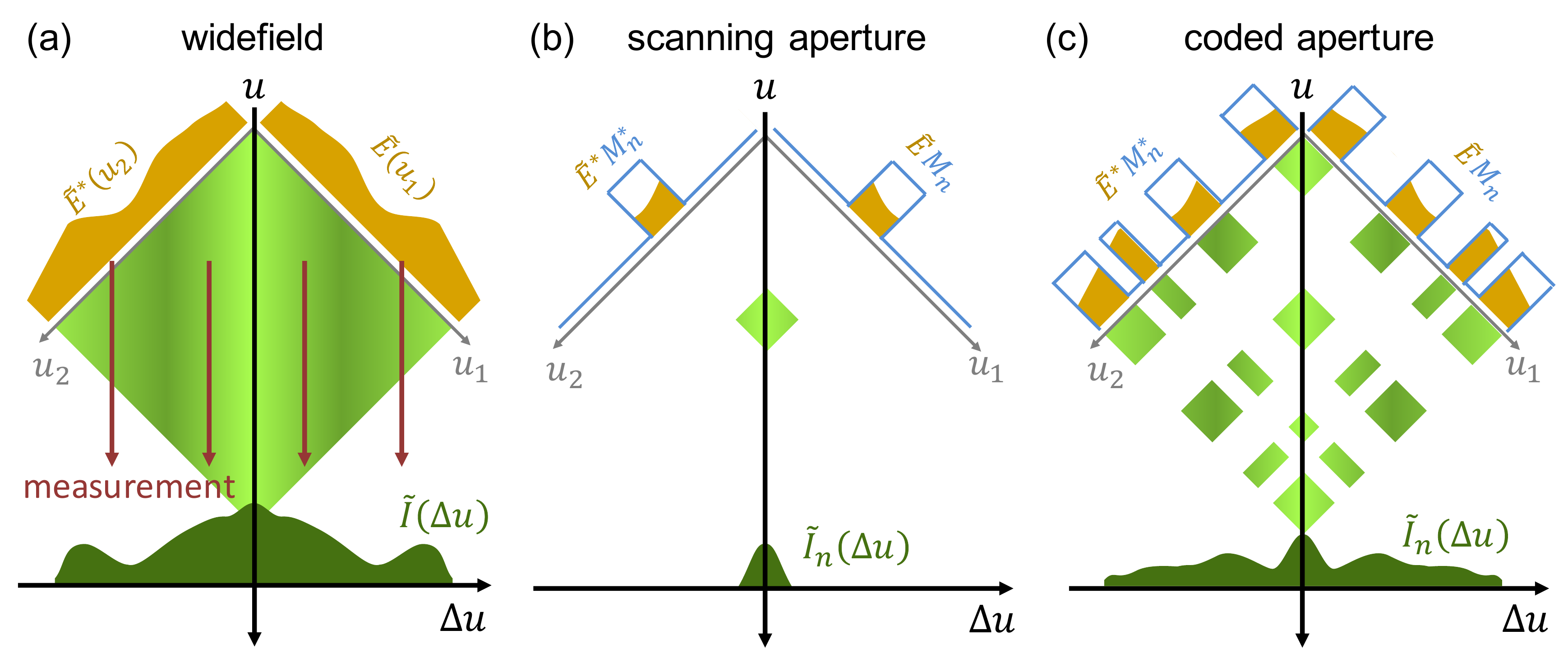}
\caption{\label{fig:multiplex-MC}Illustration of how the spatial spectrum of intensity measurements are related to the input complex-field for a coded-aperture measurement. (a) Widefield imaging: the frequency representation of the electric field $\tilde{E}(u_1)$ forms an outer product with its conjugate, resulting in a diamond-shaped Mutual Intensity (MI) space, whose rotated projection gives the spectrum of the measured image $\tilde{I}(\Delta u)$. (b) Scanning aperture imaging: the $n^{th}$ aperture $M_n$ patterns the electric field and its conjugate to probe a select area of MI space. (c) Coded-aperture imaging: the quasi-random coded apertures probe different areas within a larger range of the MI space. If $\tilde{E}$ is from a point source, the intensity projections as the source moves through focus make up the 3D OTF Fourier-transformed along the optical axis.}
\end{figure}

Our forward model relates each coded aperture's captured image to the 3D object's intensity. Each image contains information from multiple frequencies and their interference terms (which are key to resolution enhancement). The MI framework facilitates our mathematical analysis, since the projection of a mask in MI space is equivalent to applying the 3D Optical Transfer Function~(OTF) to an incoherent source (see Fig.~\ref{fig:multiplex-MC}). MI analysis further reveals the interference effects that may be obscured by looking only at a projected quantity (the OTF). For simplicity, we first describe the forward model of a point source and then generalize to multiple mutually incoherent sources.

Consider a complex electric field, specified by some properties, $\alpha$ (\textit{e.g.} location and wavelength of a point source). The field $E_s(\rbf_1;\alpha)$ acts like a unique coherent mode -- it interferes coherently with itself but not with other modes. For a single coherent mode at the front focal plane of the $4f$ system in Fig.~\ref{fig:scheme-tot}, the complex-field just before the SLM is the Fourier transform of the input complex-field~\cite{goodman05}, expressed as $\tilde{E}_s(\ubf_1,\alpha)$, where $\ubf_1$ is in units of spatial frequency that relate to the lateral coordinates $\xbf_1$ on the SLM by multiplying the wavelength $\lambda$ and the front focal length $f$, such that $\xbf_1=\lambda f \ubf_1$. The field $\tilde{E}_s(\ubf_1;\alpha)$ is then multiplied by the coded aperture, giving,
\begin{equation}
\tilde{E}_n(\ubf_1;\alpha) = M_n(\lambda f\ubf_1)\tilde{E}_s(\ubf_1;\alpha), \label{eq:modify-E0}
\end{equation}
where $\tilde{E}_n$ is the patterned complex-field in Fourier space and $M_n$ represents the $n$'th binary coded aperture pattern. At the sensor plane (the back focal plane of the $4f$ system) the intensity $I_n$ is:
\begin{align}
I_n(\rbf;\alpha) &= \iint \tilde{E}_n  (\ubf_1;\alpha) e^{ i2\pi\ubf_1\cdot\rbf}\drm^2\ubf_1 
                    \iint \tilde{E}_n^*(\ubf_2;\alpha) e^{-i2\pi\ubf_2\cdot\rbf}\drm^2\ubf_2,
\end{align}
where $\ubf_2$ is a duplicate of $\ubf_1$. The intensity image can be alternately related to the MI or WF by a simple coordinate transform $\ubf_1=\ubf+\Delta\ubf/2$, $\ubf_2=\ubf-\Delta\ubf/2$:
\begin{align}
I_n(\rbf;\alpha) &= \iiiint \tilde{E}_n(\ubf+\Delta\ubf/2;\alpha)\tilde{E}_n^*(\ubf-\Delta\ubf/2;\alpha) 
                     e^{i2\pi\Delta\ubf\cdot\rbf} \drm^2\Delta\ubf \drm^2\ubf \\
&=\iiiint \mathrm{MI}_n(\ubf,\Delta\ubf;\alpha) e^{i2\pi\Delta\ubf\cdot\rbf} \drm^2\Delta\ubf \drm^2\ubf \label{eq:image-singleMode-mi}\\
&=\iint W_n(\ubf,\rbf;\alpha) \drm^2\ubf, \label{eq:image-singleMode-wf}
\end{align}
%\begin{equation}
%\begin{split}
%W_n(\ubf,\rbf;\alpha) = \iint &M_n^*\left(\lambda f(\ubf-\frac{\Delta\ubf}2)\right) M_n\left(\lambda f(\ubf+\frac{\Delta\ubf}2)\right) \\
%&\quad\tilde{E}_s^*\left(\ubf-\frac{\Delta\ubf}2;\alpha\right)\tilde{E}_s\left(\ubf+\frac{\Delta\ubf}2;\alpha\right) e^{i2\pi\rbf\cdot\Delta\ubf} \drm^2\Delta\ubf, \label{eq:masked-wigner}
%\end{split}
%\end{equation}
where $\mathrm{MI}_n$ and $W_n$ are the MI and WF associated with the field modified by the $n$'th mask. 

Describing the intensity measurement in terms of both its WF and MI gives new insights. \eqref{eq:image-singleMode-wf} shows that the intensity image is a projection of the patterned Wigner function $W_n$ across all spatial frequencies. Alternately, looking at the Fourier transform of $I_n$, denoted by $\tilde{I}_n$, we can interpret the intensity measurement as a patterning and projection of the source MI:
\begin{align}
\tilde{I}_n(\Delta\ubf;\alpha) &= \iint \mathrm{MI}_n(\ubf,\Delta\ubf;\alpha) \drm^2\ubf \notag\\
&= \iint M_n^*\left(\lambda f(\ubf-\frac{\Delta\ubf}2)\right) M_n\left(\lambda f(\ubf+\frac{\Delta\ubf}2)\right) \mathrm{MI}_s(\ubf,\Delta\ubf';\alpha) \drm^2\ubf,\label{eq:imageFT-singleMode-mi}
\end{align}
where $\mathrm{MI}_s$ is the MI of source field $E_s$. This interpretation is illustrated for a 1D complex-field in Fig.~\ref{fig:multiplex-MC}. Each SLM pattern probes different parts of the input MI. By using multiple coded apertures, one may reconstruct the MI from its projections. 

The extension of our forward model to multiple coherent modes is straightforward. Since each mode is mutually incoherent with others, we sum over all of them with their weights $C(\alpha)$:
\begin{align}\label{eq:image-final-theo}
I_n(\rbf) &= \sum_\alpha C(\alpha)I_n(\rbf;\alpha).
\end{align}
%. The resulting formula, with $I_n(\rbf;\alpha)$ given in Eqs.~(\ref{eq:image-singleMode-mi}) and (\ref{eq:imageFT-singleMode-mi}), is

\subsection{Forward model for incoherent sources}
From the general forward model of \eqref{eq:image-final-theo}, we can now derive the case in which the object is a 3D incoherent intensity distribution (e.g. a fluorescent sample). We specify $\tilde{E}_s$ in \eqref{eq:modify-E0} to be from a single-color fluorophore located at position $(\rbf_s,z_s)$ where $z_s$ is the defocus distance. The parameter $\alpha$ is $(\rbf_s, z_s, \lambda)$ for a single point source, and our goal is to solve for the coefficients $C(\alpha)$ which represents intensity of each. To account for off-focus point sources, the field can be propagated to $-z_s$ using angular spectrum propagation~\cite{goodman05}:
\begin{equation}
\tilde{E}_s(\ubf_1;\rbf_s, z_s, \lambda) = \begin{cases}%\left\{\begin{array}{ll} 
e^{i\frac{2\pi}{\lambda}(-z_s)\sqrt{1-\lambda^2|\ubf_1|^2}-i2\pi\rbf_s\cdot\ubf_1}, & \mbox{$|\ubf_1|<\frac{NA}{\lambda}$}\\
0, & \mbox{otherwise}
\end{cases}%\end{array}\right.
.
\end{equation}
Using Eqs.~(\ref{eq:mi-def}), (\ref{eq:image-singleMode-mi}) and (\ref{eq:imageFT-singleMode-mi}), after some algebra the intensity becomes: 
\begin{equation}\label{eq:image-pointsource}
I_{n}(\rbf;\rbf_s,z_s,\lambda) = \iint K_{M_n,z_s}(\Delta\ubf;\lambda)e^{i2\pi(\rbf-\rbf_s)\cdot\Delta\ubf}\drm^2\Delta\ubf,
\end{equation}
where 
\begin{equation}\label{eq:kernel}
\begin{split}
K_{M_n,z_s}(\Delta\ubf;\lambda) = \iint &M_n^*\left(\lambda f(\ubf-\frac{\Delta\ubf}2)\right) M_n\left(\lambda f(\ubf+\frac{\Delta\ubf}2)\right)\\
&\quad e^{i\frac{2\pi z_s}{\lambda}\left(\sqrt{1-\lambda^2|\ubf-\frac{\Delta\ubf}2|^2}-\sqrt{1-\lambda^2|\ubf+\frac{\Delta\ubf}2|^2}\right)} \drm^2\ubf
\end{split}
\end{equation}
is the kernel for mask $M_n$ at depth $z_s$. Plugging \eqref{eq:image-pointsource} into \eqref{eq:image-final-theo} gives the final expression of the forward model. Before doing so, we assume that the fluorescent color spectrum $S(\lambda)$ is identical and known for all flourophores. Thus, we can decompose the mode weights in \eqref{eq:image-final-theo} into a product of the spectral weight and the 3D intensity distribution, $S(\lambda)C(\rbf_s,z_s)$, giving:
\begin{equation}\label{eq:image-final}
I_n(\rbf) = \iiint 
    \left(\iint \sum_\lambda S(\lambda) K_{M_n,z_s}(\Delta\ubf;\lambda) e^{i2\pi(\rbf-\rbf_s)\cdot\Delta\ubf} \drm^2\Delta\ubf\right) 
    C(\rbf_s,z_s)\drm^2\rbf_s \drm z_s.
\end{equation}

Equation~(\ref{eq:image-final}) describes the forward model for a 3D fluorescent object $C(\rbf_s,z_s)$ with no occlusions. The term in parentheses is a convolution kernel describing the 3D Point Spread Function (PSF) for mask $M_n$ (shown in Fig.~\ref{fig:scheme-tot}). For simplicity, we assume here no scattering, though incorporating the scattering forward model in~\cite{Liu15,pegard2016compressive} is straightforward.

\subsection{Inverse problem}
Based on the raw data and forward model, the inverse problem is formulated as a nonlinear optimization. Our goal is to reconstruct the 3D intensity distribution $C(\rbf_s,z_s)$ from the measured images. To do so, we aim to minimize data mismatch, with an $\ell_1$ regularizer to mitigate the effects of noise (and promote sparsity where applicable). The mismatch is defined as the least-square error between the measured intensity images and the intensity predicted by our forward model (Eq.~(\ref{eq:image-final})). This formulation has a smooth part and a non-smooth part in the objective function and is efficiently solved by a proximal gradient descent solver (FISTA~\cite{beck2009fista}).

To formulate the inverse problem, we first discretize the forward model in \eqref{eq:image-final} to be
\begin{equation}
\ybm = \Abm\cbm.
\end{equation}
Here $\ybm\in \mathbb{R}^{MP\times1}$ corresponds to predicted images on the sensor; each small chunk ($\in \mathbb{R}^{P\times1}$) of $\ybm$ is a vectorized image $I_n(\rbf)$. We discretize $\rbf$ into $P$ pixels, and the number of masks is $M$, so $n=1\ldots M$. Similarly, we discretize $\rbf_s$ and $z_s$ into $P'$ pixels and $L$ samples, respectively, to obtain a vectorized version of $C(\rbf_s,z_s)$ ($\cbm\in \mathbb{R}^{LP'\times 1}$). The matrix $\Abm\in \mathbb{R}^{MP\times LP'}$, which is not materialized, represents the summation and convolution in \eqref{eq:image-final} using 2D Fast Fourier Transforms (FFTs) for each subvector ($\in \mathbb{R}^{P'\times1}$) of $\cbm$, with zero-padding to avoid periodic boundary condition errors. The convolution kernel is precomputed and stored for speed.

The inverse problem becomes a data fidelity term plus an $\ell_1$ regularization parameter $\mu$:
\begin{equation}\label{eq:optimization-discre}
\argmin_{\cbm\geq0} \frac{1}{2}\|\ybm-\ybm_{\mathrm{meas}}\|_2^2 + \mu\|\Wbm\cbm\|_1,
\end{equation}
where $\ybm_{\mathrm{meas}}\in \mathbb{R}^{MP\times1}$ is the measured intensity. We also use a diagonal matrix $\Wbm\in \mathbb{R}^{LP'\times LP'}$ to lower the weight of point sources near the borders of images whose light falls off the sensor. Each diagonal entry of $\Wbm$ is obtained by summing the corresponding column in $\Abm$. Outside point sources may also contribute to the measured intensity due to defocus; hence, we use an extended field-of-view method~\cite{bertero2005} to solve for more sample points in $\cbm$ than $\ybm$ (\textit{i.e.} $P'>P$).

\section{Design of coded apertures}\label{sec:maskdesign}
\begin{figure}
\centering
\includegraphics[width=0.65\textwidth]{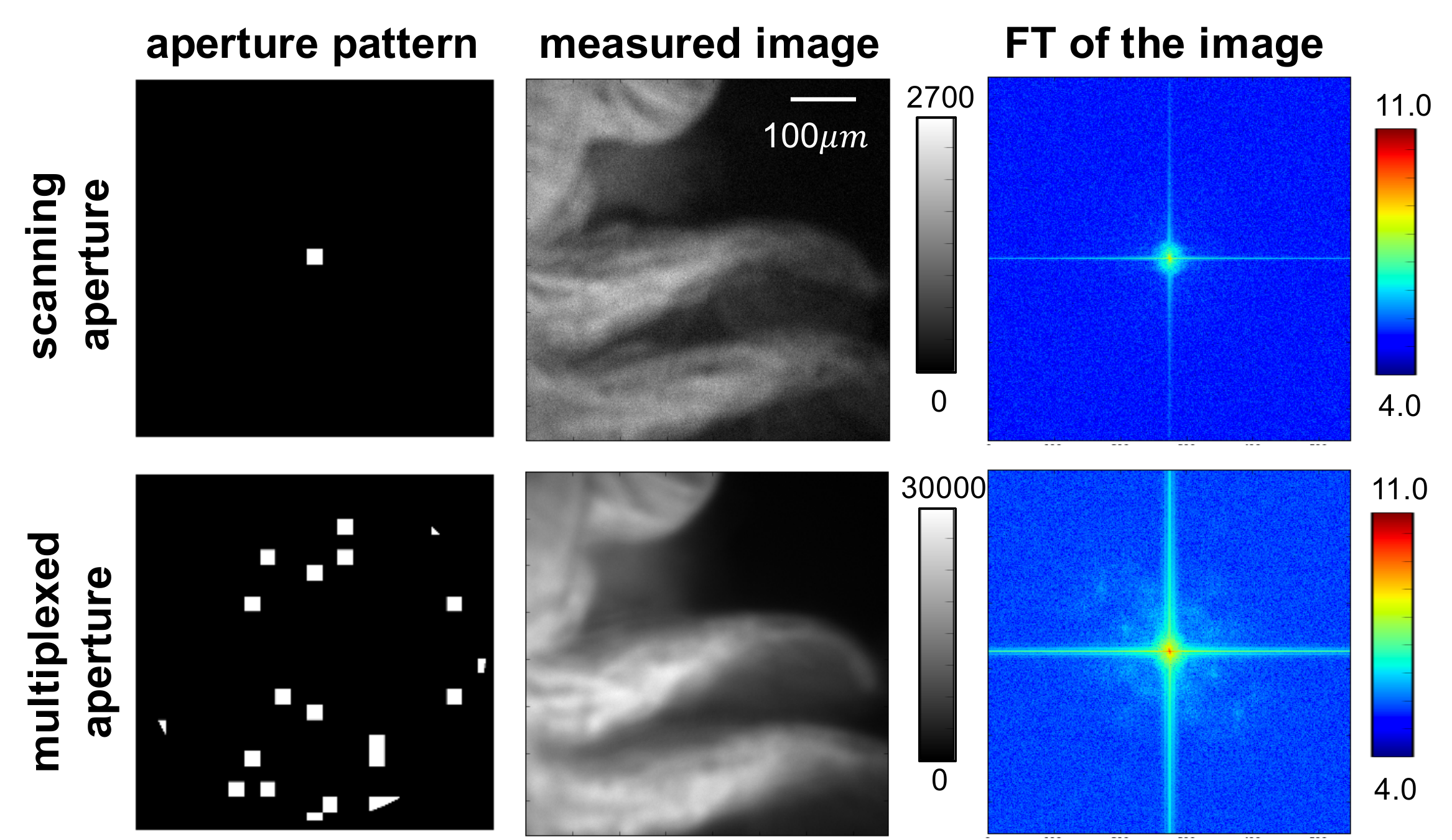}
\caption{\label{fig:comp-res}Multiplexed phase-space measurements contain more information than scanning-aperture measurements. (Left) A sample aperture pattern, (Middle) the corresponding measured intensity image (same exposure time), and (Right) its log-scale Fourier transform. The multiplexed measurements have better light throughput and more high-frequency content.}
\end{figure}

In the scanning-aperture scheme~\cite{waller12NP,Liu15}, smaller apertures give better frequency sampling of the 4D phase space, at a cost of: 1) lower resolution, 2) lower signal-to-noise ratio (SNR) and 3) large data sets. Our multiplexing scheme alleviates all of these problems. Multiplexing achieves diffraction-limited resolution by additionally capturing interference terms, which cover the full NA-limited bandwidth. This is evident in the Fourier transform of the captured images (Fig.~\ref{fig:comp-res}). The SNR improvement is also visible; the multiplexed image is less noisy. 

Our masks are chosen by quasi-random non-replacement selection. We section the SLM plane into 18$\times$18 square blocks and keep only the 240 blocks that are inside the system NA. For each mask, we open 12 blocks, selected randomly from the blocks remaining after excluding ones that were open in previous sequences. In this scheme, the full NA can be covered by 20 masks. To allow for both diversity and redundancy, we choose to to cover the entire pupil 5 times, resulting in 100 multiplexed aperture patterns, one of which is shown in Fig.~\ref{fig:comp-res}.

Importantly, the number of multiplexed patterns can be flexibly chosen to trade off accuracy for speed of capture. For instance, by increasing the number of openings in each pattern, we can cover the entire pupil with fewer patterns. This means that we may be able to reconstruct the object from fewer measurements, if the inverse problem is solvable. By using \textit{a priori} information about the object (such as sparsity in 3D) as a constraint, we can solve under-determined problems with fewer total measured pixels than voxels in the reconstruction.

\section{Experiments}\label{sec:exp}
%\subsection{Experimental Setup}\label{sec:exp-setup}
%\begin{figure}
%\centering
%\includegraphics[width=\textwidth]{fig/setup2.pdf}
%\caption{\label{fig:setup}(a) The experimental setup scheme containing additional relay optics. $P_2$ is relayed to $P_4$ and an SLM is put on $P_4$. (b) A picture of the real setup, with an extra polarization beam splitter because the SLM is polarization dependent.}
%\end{figure}
Our experimental setup consists of the $4f$ system ($f_1=250~mm$,$f_2=225~mm$) shown in Fig.~\ref{fig:scheme-tot}, with an additional 4$f$ system in front, made of an objective lens (20$\times$ NA 0.4) and a tube lens ($f=200~mm$) to image the sample at the input plane. The SLM (1400$\times$1050 pixels of size $10.3\mum$) is a liquid crystal chip from a 3-LCOS projector (Canon SX50) which is reflective and polarization-sensitive, so we fold the optical train with a polarization beam splitter and insert linear polarizers. Our sensor (Hamamatsu ORCA-Flash4.0 V2) captures the multiplexed images and synchronizes with the SLM via computer control.

%\subsection{Results and Discussion}
\begin{figure}[tb]
\centering
\includegraphics[width=\textwidth]{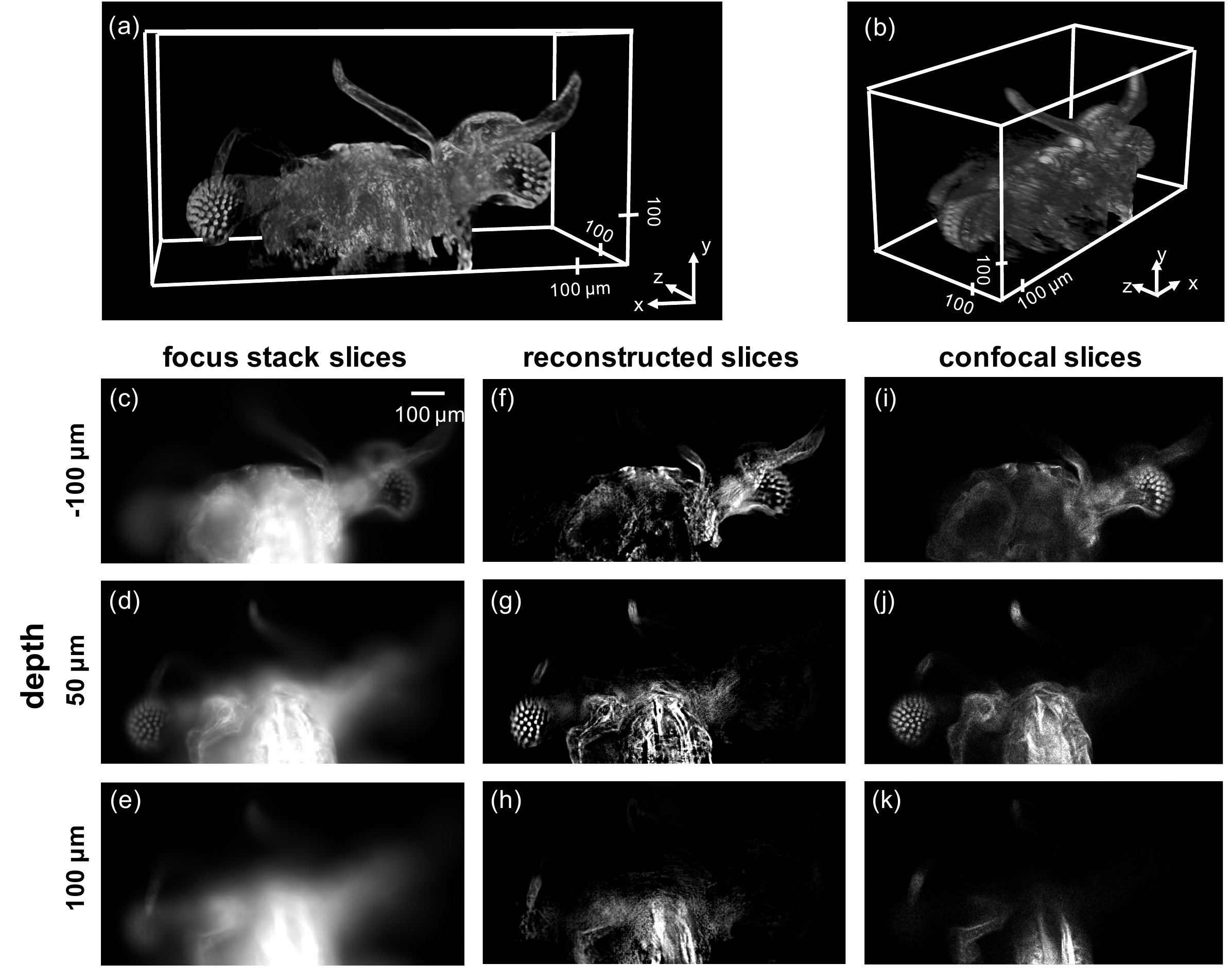}\\[10pt]
\caption{\label{fig:rec-3d}3D reconstruction of a brine shrimp sample as compared to focus stack and confocal microscopy. (a) and (b) 3D renderings of the reconstructed fluorescence intensity distribution (1010$\times$510$\times$500$\mum$) from different perspectives. (c)-(e) 2D widefield images at three different focus planes. (f)-(h) Slices of our reconstructed volume at the same depth planes. (i)-(k) Confocal microscopy slices at the same depth planes for comparison.}
\end{figure}

Our sample is a fixed fluorescent brine shrimp (Carolina Biological). It is relatively dense, yet does not fill the entire 3D volume. The reconstructed 3D intensity (Fig.~\ref{fig:rec-3d}(a), \ref{fig:rec-3d}(b) and \ref{fig:rec-3d}(f)-\ref{fig:rec-3d}(h)) is stitched from five volume reconstructions, each with 640$\times$640$\times$120 voxels to represent the sample volume of 455$\times$455$\times$600$~\mum$. The reconstruction is cropped to the central part of our extended field of view, so the final volume contains 1422$\times$715$\times$100 voxels corresponding to 1010$\times$510$\times$500$\mum$. The dataset size is large (9~GB), and since the size of the 3D array is $\sim$5$\times10^7$ without the extended field-of-view and the measured data is $\sim$4$\times10^7$, the number of operations for evaluating Eq.~(\ref{eq:image-final}) is on the order of 3$\times10^{10}$. This takes 4 seconds to compute on a computer with 48-core 3.0~GHz CPUs and requires 94~GB memory to store the kernel (\eqref{eq:kernel}). 

The reconstructed 3D intensity is shown in Fig.~\ref{fig:rec-3d}, alongside images from a confocal microscope and a widefield focus stack, for comparison. Both our method and the focus stack use a 0.4 NA objective, while the confocal uses 0.25 NA; hence, the confocal results should have slightly better resolution. Our reconstructed slices appear to have slightly lower resolution than the defocus stack and confocal, possibly due to the missing information in the frequency mutual intensity illustrated in Fig.~\ref{fig:multiplex-MC}(c), which is greatly undersampled. As expected, the depth slices of our reconstruction have better rejection of information from other depths, similar to the confocal images.

\begin{figure}
\centering
\includegraphics[width=0.98\textwidth]{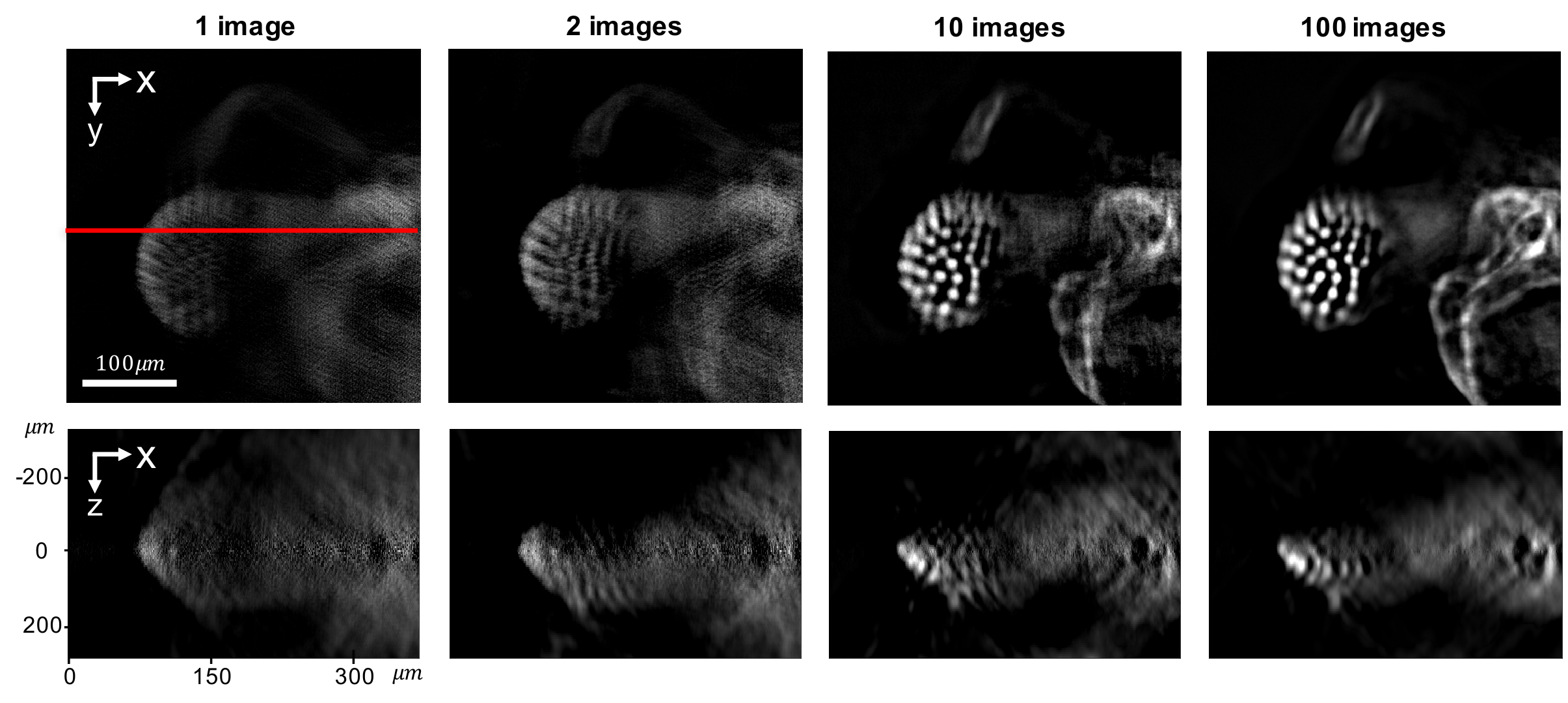}
\caption{\label{fig:subsample-comp}Image quality can be traded for capture speed (number of coded aperture images). 3D reconstructions from increasing numbers of images with different coded apertures show that this object is too dense to be accurately reconstructed by a single coded-aperture image, but gives a reasonable reconstruction with 10 or more images, due to sparsity of the sample.}
\end{figure}

To illustrate the flexible tradeoff between capture time (number of coded apertures used) and quality, we show reconstructions in Fig.~\ref{fig:subsample-comp} using different numbers of coded aperture images. The case of only 1 image corresponds to a single coded aperture and gives a poor result, since the sample is relatively dense. However, with as few as 10 images we obtain a reasonable result, despite the fact that we are solving a severely under-determined problem. This is possible because the measurements are multiplexed and so the $\ell_1$ regularizer acts as a sparsity promoter.

\section{Conclusion}
We demonstrated 3D reconstruction of a large-volume high-resolution dense fluorescent object from multiplexed phase-space measurements. An SLM in Fourier space dynamically implements quasi-random coded apertures while intensity images are collected in real space for each coded aperture. Theory is developed in the framework phase space, with relation to Mutual Intensity functions and 3D OTFs. Reconstruction is formulated as an $\ell_1$-regularized least-square problem. This method enables diffraction-limited 3D imaging with high resolution across large volumes, efficient data capture and a flexible acquisition scheme for different types and sizes of samples.

\section*{Funding}
The Office of Naval Research (ONR) (Grant N00014-14-1-0083).
% National Science Foundation (NSF) (1253236, 0868895, 1222301); Program 973 (2014AA014402); Natural National Science Foundation (NSFC) (123456).

\section*{ACKNOWLEDGMENTS}
The authors thank Eric Jonas, Ben Recht, Jingzhao Zhang and the AMP Lab at UC Berkeley for help with computational resources. 
\end{document}